\documentclass[%
twocolumn,
10pt,
nofootinbib,
amsmath,amssymb,
aps,
prd,
showkeys
]{revtex4-2}
\usepackage[english]{babel}
\usepackage{newtxtext,newtxmath}
\usepackage{graphicx}\graphicspath{{figures/}}
\usepackage{dcolumn}  
\usepackage{bm}  
\usepackage{xcolor}\definecolor{darkblue}{rgb}{0,0,0.5}\definecolor{firebrick}{rgb}{0.75,0.125,0.125}\definecolor{darkgreen}{rgb}{0,0.45,0}
\usepackage[colorlinks=true,linkcolor=firebrick,citecolor=darkgreen,urlcolor=darkblue]{hyperref}
\usepackage[capitalise]{cleveref}
\usepackage{wrapfig}
\usepackage{gensymb}
\usepackage{siunitx}
\usepackage[T5,T1]{fontenc}
\usepackage[utf8]{inputenc}
\usepackage{enumitem}
\usepackage{balance}
\usepackage{soul}

\newcounter{appendixref}
\crefname{appendixref}{Appendix}{Appendices}
\crefformat{appendixref}{#2Appendix#3}

\begin{document}

\title{Evidence for the Moon and Sun shadows with cosmic rays at an average energy of $\text{7}{\times}\text{10}^\text{17}$\,eV}

\author{The Pierre Auger Collaboration}
\email{spokespersons@auger.org}
\affiliation{Pierre Auger Observatory, Av.\ San Mart\'in Norte 306, 5613 Malarg\"ue, Mendoza, Argentina}
\thanks{http://www.auger.org}

\date{\today}  

\begin{abstract}
Interactions of cosmic rays with the Moon and the Sun produce deficits in their arrival-direction distributions relative to an isotropic flux.
Such shadows have been observed previously at energies between \qty{e12}{eV} and \qty{e16}{eV}.
We report the first evidence for the Moon and Sun shadows at cosmic-ray energies larger than about \qty{e16}{eV} (average energy of \qty{7e17}{eV}), using data collected by the Pierre Auger Observatory.
We employ data from three detector arrays covering \qty{3000}{km^2}, \qty{27}{km^2}, and \qty{2}{km^2}, with spacings of \qty{1500}{m}, \qty{750}{m}, and \qty{433}{m}, respectively.
The data amount to over 10.6 million events.
The Moon and Sun shadows are detected with a combined significance of approximately $3\sigma$.
These observations confirm the pointing accuracy of the Surface Detector of the Pierre Auger Observatory using celestial bodies.
From the combined Sun and Moon shadows, we infer an overall angular resolution of $(0.59^{+0.15}_{-0.11})\degree$.
\end{abstract}


\maketitle

\section{Introduction}

The shielding of cosmic rays by the Moon and the Sun was first predicted in 1957 by Clark~\cite{Clark57}.
Characterizing these shadows in data is a standard technique for validating the angular resolution and pointing accuracy of astroparticle observatories.
For an ideal detector, the deficit from an isotropic cosmic-ray flux would form a disk-shaped shadow with an angular radius corresponding to that of the Moon and the Sun, $\delta_\text{c}\approx \ang{0.26}$.
Finite detector resolution causes events to migrate into the shadow, and the resulting blurring can be used to infer the angular resolution of an observatory.

Previous cosmic-ray measurements of these shadows were limited to energies up to the PeV scale~\cite{PhysRevD.43.1735_Cygnus,1991ICRC....2..708A_EASTOP,1993ICRC....4..351A_Tibet,PhysRevD.47.2675,PhysRevD.49.1171_CASA,PhysRevD.61.092002,Amenomori_2000,2001ICRC....2..594S_Milagro,2003ICRC....7.4065X,AMBROSIO2003145,Achard_2005,ADAMSON2011457,PhysRevD.84.022003_ARGO,PhysRevLett.111.011101,Aartsen_2014_icecube,bird2015observingcosmicraymoon,PhysRevD.97.102005,PhysRevD.98.123011,PhysRevD.98.123012,Aartsen_2019,Nan:2021OF,Grapes_PhysRevD.106.022009,Alfaro_2024}, where the flux provides sufficient statistics.
At energies above \qty{e16}{eV}, the steeply falling cosmic-ray spectrum makes such observations increasingly challenging.

Using data from the Pierre Auger Observatory, we present the first observation of the Moon and Sun shadows at an average energy of \qty{7e17}{eV}.

The Observatory~\cite{2015172_NIM_Auger}, located in the Pampa Amarilla region of the Province of Mendoza, Argentina, comprises a Surface Detector (SD) array and a Fluorescence Detector (FD).
The SD consists of more than 1600 water-Cherenkov detectors deployed on three nested grids with spacings of \qty{1500}{m} (SD-1500), \qty{750}{m} (SD-750), and \qty{433}{m} (SD-433).
The SD-1500 array covers an area of \qty{3000}{km^2}, the SD-750 array \qty{27}{km^2}, and the SD-433 array \qty{2}{km^2}.
The different detector spacings and collection areas enable measurements of cosmic rays over a wide energy range, from about \qty{e16}{eV} up to above \qty{e20}{eV}.
Although the cosmic-ray rate in this energy range is modest (a few events per minute), the combination of data from the three arrays, which have operated stably for extended periods (almost 20 years for SD-1500, 15 years for SD-750, and 3 years for SD-433), provides the exposure required to probe the Moon and Sun shadows.
With more than 10 million events, the combined data set enables the first observation of these shadows at an average energy of \qty{7e17}{eV}, extending such measurements by nearly three orders of magnitude beyond previous studies.
Moreover, the angular resolutions of the three arrays, discussed in Refs.~\cite{Aab_2020_Auger_reco,HAS_reco_2014,AMIGA_2011,Silli:2021Jt_SD-433}, are adequate to observe these shadows.

Although the primary scientific focus of the Observatory is the study of (charged) cosmic rays, which are deflected by magnetic fields so that their arrival directions at Earth are offset from the position of their sources by angles much larger than the reconstruction uncertainty, the validation of the angular resolution and pointing is essential in the broader context of the Observatory's role in multimessenger astronomy.
In particular, the Pierre Auger Collaboration has been conducting searches for neutral particles such as neutrons~\cite{neutron_search}, neutrinos~\cite{neutrinos_2013}, and gamma rays~\cite{photon_searches}, which would point directly back to their sources, and has participated in follow-up observations of gravitational-wave events~\cite{GW_followup_nu_PhysRevD.94.122007,GW_followup_photons}.

We first present the data sets (\cref{s:data}), then we detail the measurement of the flux deficit due to the absorption of cosmic rays by the Moon and the Sun (\cref{sec:cumul_plot}), and we finally obtain the estimate of the angular resolution (\cref{s:angular}).
Our conclusions are given in \cref{s:conclusion}.

\section{The data}
\label{s:data}

A detailed description of the water-Cherenkov detectors, which are the building blocks of the three SD arrays exploited here, has been published in Ref.~\cite{2015172_NIM_Auger}.
Here we provide only brief remarks concerning features relevant to the data used in this work.
Each detector is equipped with three 9-inch photomultiplier tubes (PMTs) to record the Cherenkov light produced by secondary particles from extensive air showers reaching the ground.
The signals are digitized using a 10-bit FADC readout.
Each detector station operates autonomously: it is powered by solar panels, it is synchronized using GPS timing, and it transmits data to a central data acquisition system via radio.
An SD event is recorded when at least three stations register time-coincident signals.
The data used in this analysis represent the Phase I data set and were recorded between January 2004 and March 2023, from the start of data taking of each array until the data recorded with the upgraded electronics~\cite{2016arXiv160403637T}.

The reconstruction procedure is only applied to events in which the SD station with the largest signal is surrounded by at least 5 active stations.
The arrival direction of each event is reconstructed by fitting the air-shower front to the signal start times recorded by the detector stations.
An initial direction estimate is obtained assuming a planar shower front and is subsequently refined using a more realistic spherical-front model.
The energy of each event is obtained from the calibration of the energy estimator, defined for each array as the reconstructed signal at a specific distance from the air-shower axis, cross-calibrated with the energy obtained from the FD~\cite{spectrum_across_declinations,PhysRevD.102.062005,JCAP_HAS_2015,Abreu_EPJC2021_spectrum_sd750,BrichettoOrquera:202340}.
Further details on the event reconstruction are provided in Refs.~\cite{Aab_2020_Auger_reco,HAS_reco_2014,Silli:2021Jt_SD-433}.
Previous studies of the angular resolution of the three arrays reported an average resolution of about \ang{1} for events with low station multiplicities, improving to better than \ang{0.4} when more than 7 stations participate in the reconstruction~\cite{Aab_2020_Auger_reco,neutron_icrc23,AMIGA_2011,Silli:2021Jt_SD-433}. The same arrival-direction reconstruction algorithms are used for all arrays, and since angular resolution is primarily determined by the number of stations rather than the spacing of the array, the resulting data from the different arrays are similar and can be combined.

\begin{figure}
    \includegraphics[width=\linewidth]{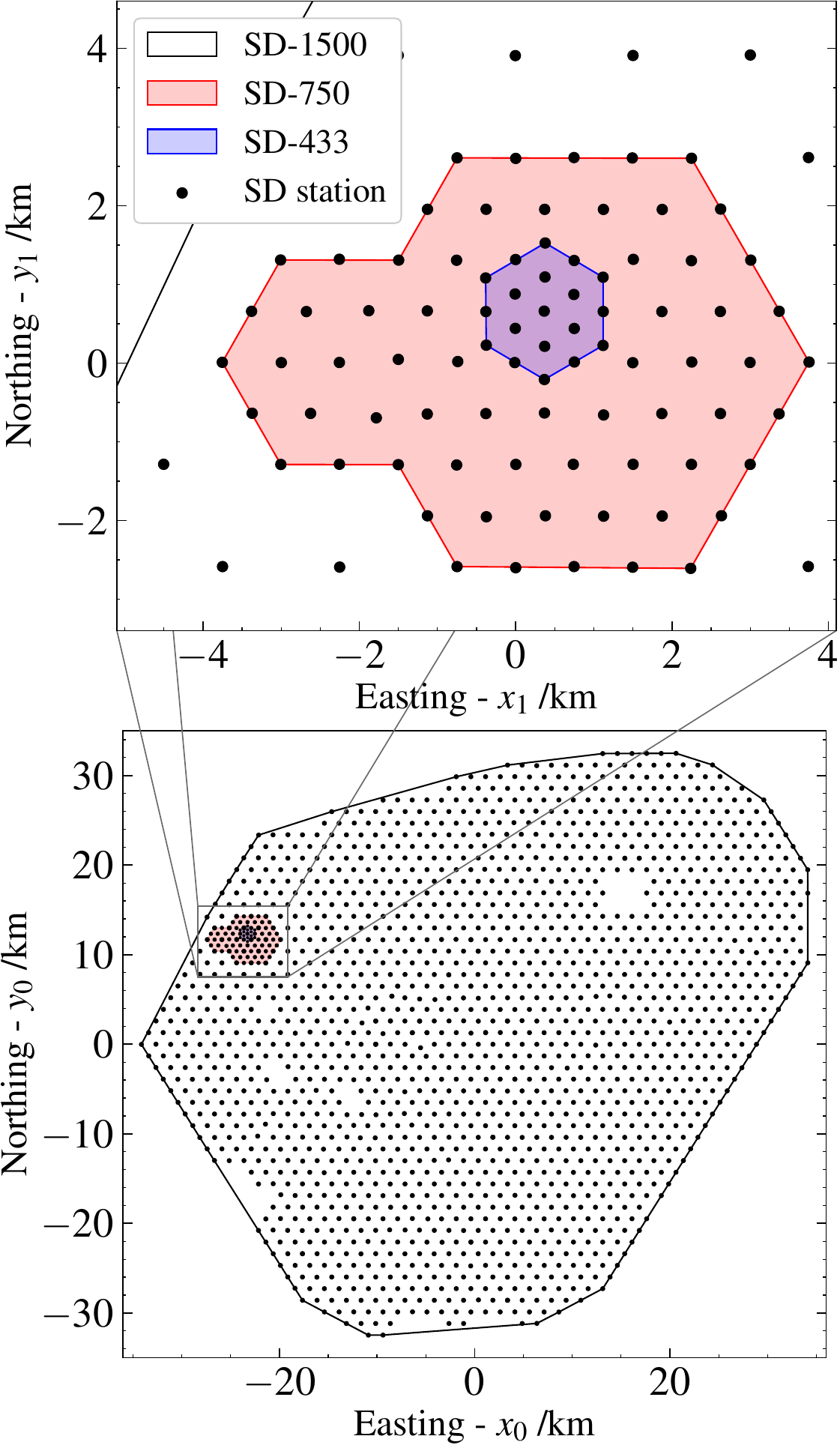}
    \caption{The configuration of the SD arrays. The top panel displays the SD-433 array (blue) and the SD-750 array (red), while the bottom panel displays the SD-1500 array. Black dots represent the locations of the individual detector stations.
    The areas have been centered on the middle of the arrays, with UTM coordinates $(x_0,y_0)=(474255, 6102240)$\,m and $(x_1,y_1)=(450627, 6113915)$\,m.}
    \label{fig:array}
\end{figure}

The total dataset comprises 10.6 million events, obtained by including all reconstructed events, independently of their energy or zenith angles.
The number of events in each data subset is listed in \cref{tab:allEvents}.
Since the arrays are nested, as illustrated in \cref{fig:array}, an event can be recorded by more than one array: if this is the case, only the reconstruction from the array with the smallest detector spacing is retained.
This ensures that more stations are selected, leading to a better reconstruction. Events recorded by the SD-1500 array are subdivided according to the reconstruction algorithm into vertical ($\theta\leq\ang{65}$) and inclined ($\ang{60}<\theta<\ang{80}$) air showers.
In the overlapping zenith-angle range, the vertical reconstruction is prioritized.

The event numbers are dominated by the SD-750 array, which contributes 64\% of the events, followed by the SD-1500 and SD-433 arrays.
Each array operates with its own trigger based on similar time-coincidence algorithms optimized for the respective detector spacings.\footnote{For the SD-750 data, the station-level trigger is enhanced to allow for increased sensitivity to lower signals.}
The unequal event fractions among the arrays arise from their different areas, detector spacings, and operational periods.
The data taking began in 2004 for SD-1500, 2008 for SD-750, and 2018 for the SD-433 array (see \cref{tab:allEvents}).

\begin{table}
\caption{Number of events in each data set, fraction, $f$, the average number of stations participating in the reconstruction, $\langle n\rangle$, and the data-taking period.
Events recorded by SD-1500 are separated into vertical (vert.)\ and inclined (incl.)\ categories based on the reconstruction.}
\label{tab:allEvents}
\begin{ruledtabular}
\begin{tabular}{lrccr}
Type & Events & $f/\%$ & $\langle n\rangle$ & Time Period
\\\hline
SD-1500 vert. & \num{2537224} & \num{23.8} & \num{4.0} & Jan '04--Dec '22
\\
SD-1500 incl. & \num{592455} & \phantom{0}\num{5.6} & \num{5.9} & Jan '04--Mar '23
\\
SD-750 & \num{6828339} & \num{64.1} & \num{3.6} & Jan '08--Feb '23
\\
SD-433 & \num{694755} & \phantom{0}\num{6.5} & \num{5.0} & Jan '18--Dec '21
\\\hline
Total & \num{10652773} & 100 & \num{3.9} &
\end{tabular}
\end{ruledtabular}
\end{table}

\begin{figure}
\includegraphics[width=\columnwidth]{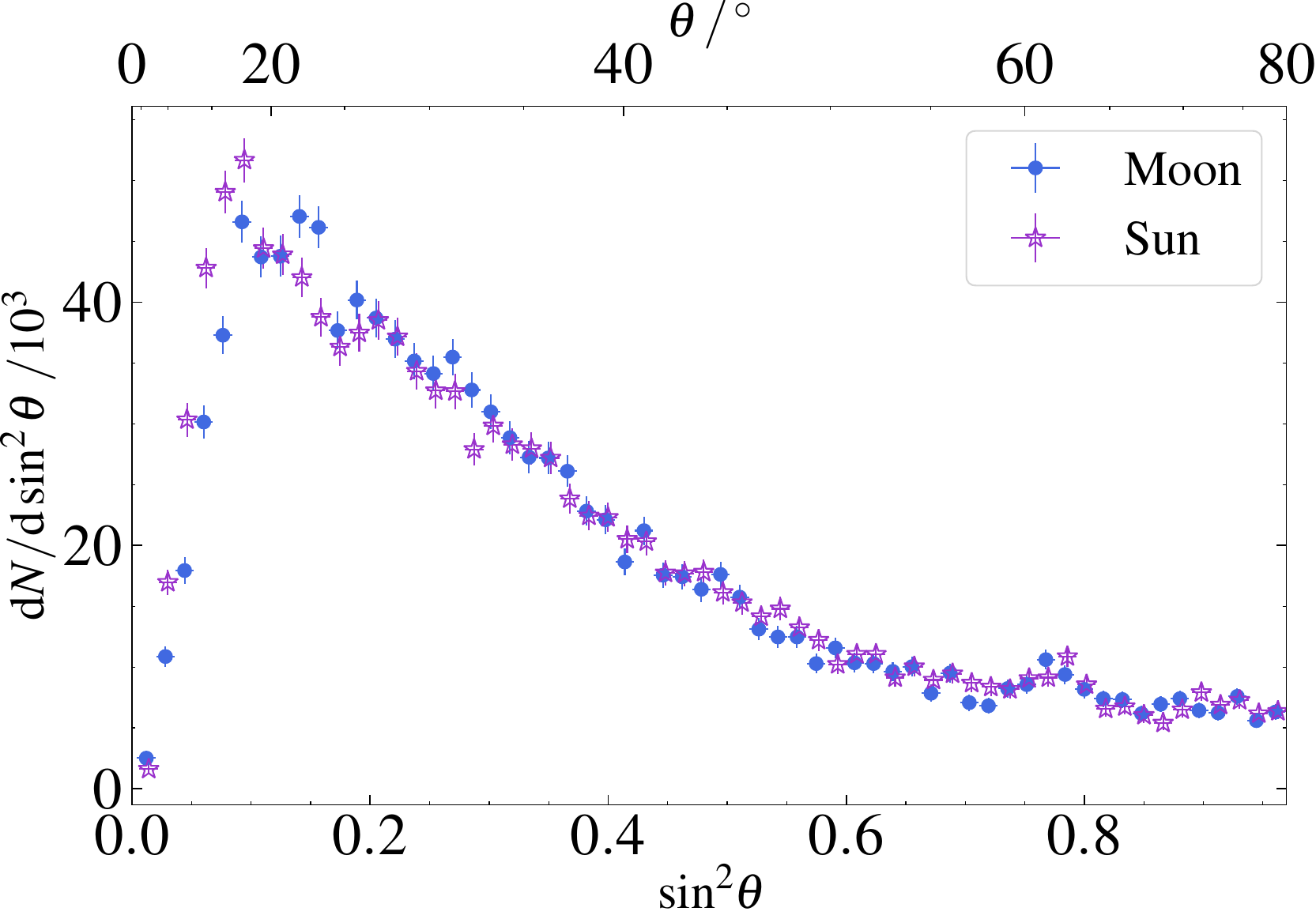}
\caption{Distribution of $\sin^2\theta$ for the zenith angles, $\theta$, of the events within \ang{5} of the Moon (filled blue circles) and the Sun (empty purple stars).}
\label{fig:zenith_moon_sun}
\end{figure}

\begin{table*}
\caption{Summary of events arriving within \ang{5} of the Moon and Sun directions.
Events recorded by SD-1500 are separated into vertical (vert.) and inclined (incl.) categories based on the reconstruction.}
\label{tab:within5deg}
\begin{ruledtabular}
\begin{tabular}{lrr rrr rrr rr}
Type & \multicolumn{2}{c}{Events} & \multicolumn{2}{c}{Fraction (\%)} & \multicolumn{2}{c}{Median $\theta/^\circ$} & \multicolumn{2}{c}{Avg.\ Multiplicity} & \multicolumn{2}{c}{$\langle E\rangle/\text{EeV}$}
\\
& Moon & Sun & Moon & Sun & Moon & Sun & Moon & Sun & Moon & Sun
\\\hline
SD-1500 vert. & \num{4647} & \num{4960} & \num{26} & \num{27} & \num{40.3} & \num{40.8} & \num{4.1} & \num{4.1} & \num{1.94} & \num{1.98}
\\
SD-1500 incl. & \num{1094} & \num{1104} & \num{6} & \num{6} & \num{69.8} & \num{70.3} & \num{5.9} & \num{5.9} & \num{1.56} & \num{1.53}
\\
SD-750 & \num{11340} & \num{11538} & \num{62} & \num{62} & \num{28.8} & \num{27.8} & \num{3.7} & \num{3.6} & \num{0.11} & \num{0.10}
\\
SD-433 & \num{1134} & \num{1048} & \num{6} & \num{6} & \num{26.4} & \num{26.7} & \num{5.1} & \num{5.1} & \num{0.05} & \num{0.05}
\\\hline
Total & \num{18215} & \num{18650} & 100 & 100 & \num{31.9} & \num{31.4} & \num{4.0} & \num{4.0} & \num{0.66} & \num{0.68}
\end{tabular}
\end{ruledtabular}
\end{table*}

The reconstruction algorithms provide the arrival direction of each event, which is compared to the astrometric positions of the Moon and Sun.
The positions of the two celestial bodies at the time of each event are determined using the \texttt{Astropy} Python library~\cite{astropy:2022}.
The angular separation (great-circle angular distance) $\delta$ between the reconstructed arrival direction and the center of Moon or Sun is then calculated for all events, out of which we select events with $\delta<\ang{5}$.
The selected event counts for the Moon, \num{18215}, and for the Sun, \num{18650}, are similar.
The numbers of events contributed by the individual arrays are summarized in \cref{tab:within5deg}, together with their median zenith angles and their average station multiplicities.
The zenith-angle distributions of these events for both the Moon and the Sun are shown in \cref{fig:zenith_moon_sun}.
The two distributions of event counts near the two celestial bodies are statistically compatible, except at $\theta<\ang{18}$ (or $\sin^2\theta<0.1$), where the number of events associated with the Sun is larger.
The difference is attributed to the precession of the Moon's orbit, which is tilted \ang{5} with respect to the ecliptic and precesses with a period of 18.6 years.
By chance, the inclination relative to Earth's equator was minimum in October of 2016, so the Moon did not reach altitude angles as high as the Sun's highest during half of the precession period, between 2012 and 2021, constituting the bulk of our data.

\section{The shadows of the Moon and the Sun}
\label{sec:cumul_plot}

To measure the flux deficit caused by occultation of cosmic rays by the Moon and the Sun, we use the expectation for the distribution of the angular distance $\delta$ in the absence of such occultation.
This expectation is based on a constant event count per unit solid angle (sr), $(\mathrm{d}N/\mathrm{d}\Omega)_\text{exp}$, in a small region around the celestial bodies.
The uniformity of $(\mathrm{d}N/\mathrm{d}\Omega)_\text{exp}$ has been verified using two shuffling techniques.
By shuffling either the time or the azimuth angle of the events, we determined the background as a function of $\delta$, which showed a constant behavior.
We determine it separately for each celestial body as the number of events with $\delta<\ang{5}$ divided by the solid angle corresponding to the region.
The solid angle subtended by the Moon or the Sun shadow has been subtracted from the $\delta<\ang{5}$ total solid angle.
We find $(\mathrm{d}N/\mathrm{d}\Omega)_\text{exp}=\qty{764000}{\per\steradian}$ for the region around the Moon and $(\mathrm{d}N/\mathrm{d}\Omega)_\text{exp}=\qty{782000}{\per\steradian}$ for the region around the Sun.
The statistical uncertainty, $\varepsilon_\text{exp}$, of the uniform density is due to the Poisson uncertainty in the number of events with $\delta<\ang{5}$ and amounts to ${\sim}\qty{6000}{\per\steradian}$ (0.7\%) in both cases.

\begin{figure}
\includegraphics[width=\columnwidth]{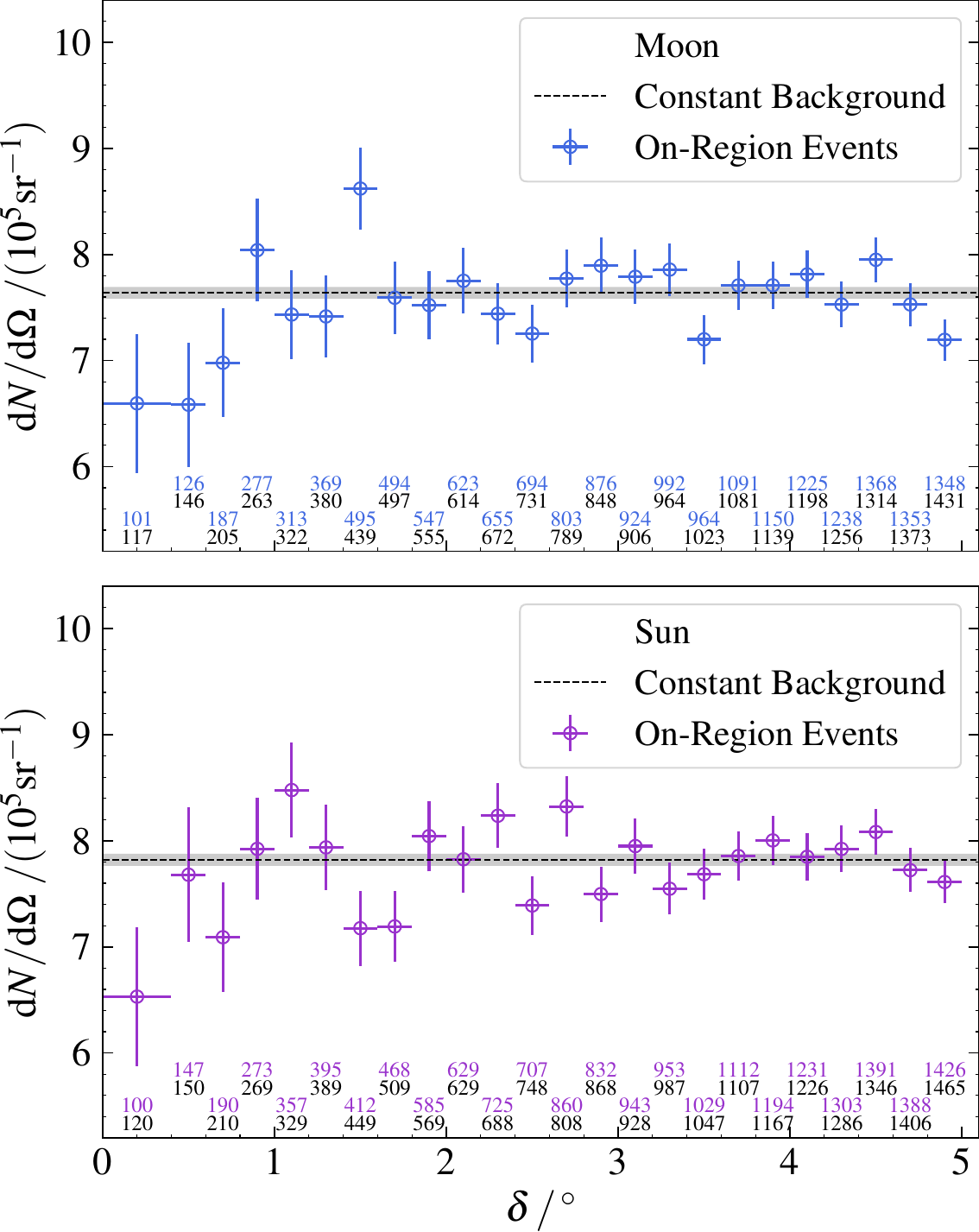}
\caption{The event count per solid angle as a function of $\delta$ for the Moon (upper panel) and the Sun (lower panel).
The uniform expectation is shown as the horizontal black line with the band representing the statistical uncertainty.
Note that the first two bins are merged to reduce the statistical uncertainties.
The measured number of events in each bin is written in color, while the unblocked expectation is in black.
The statistical uncertainty of the measured numbers decreases from 10\% to 3\% with increasing $\delta$.
For the uniform expectation, the statistical uncertainty is 0.7\%.}
\label{fig:binned_moon_sun}
\end{figure}

The number of events per sr in the Moon- and Sun-centered regions is binned in equal $\delta$-intervals of \ang{0.2}, and compared to the uniform background expectation in \cref{fig:binned_moon_sun}.
The first two bins have been merged to increase the statistics.
For both the Moon and Sun, a deficit of events can be observed in the first three bins, corresponding to angular distances $\delta<\ang{0.8}$, although this deficit is less pronounced for the Sun.

From the measured uniform densities and the solid angles of the Sun and the Moon, the expected number of occluded events by the Moon is 49.4, and by the Sun, 50.6.
For both the Moon and Sun a deficit of 54 and 43 events, respectively, can be observed in the first three bins, corresponding to angular distances $\delta<\ang{0.8}$, where the deficit is less pronounced for the Sun.

To better represent the deficit, we transition from binned event counts to cumulative distributions, which provide improved sensitivity to localized deficits.
To assess the significance of the deficit, we compare the cumulative number of observed events $N_\text{obs}(\delta)$ with an angular distance smaller than $\delta$ from the center of the Moon (or Sun) to the corresponding uniform expectation 
\begin{align}
  N_\text{exp}(\delta ) =
    \Omega_\delta \left(\frac{\mathrm{d}N}{\mathrm{d}\Omega}\right)_\text{exp},
\end{align}
where $\Omega_\delta=2\pi(1-\cos\delta)$ is the solid angle of radius $\delta$.
We define the relative difference, along with its uncertainty, as
\begin{align}
  \Delta(\delta) =
    \frac{N_\text{obs}(\delta)}{N_\text{exp}(\delta)} - 1 \pm 
    \frac{N_\text{obs}(\delta)}{N_\text{exp}(\delta)}
    \sqrt{\frac{1}{N_\text{obs}(\delta)} +
          \frac{\varepsilon_\text{exp}^2\Omega_\delta^2}{N_\text{exp}(\delta)^2}
    }.
\end{align}

\begin{figure}
\includegraphics[width=\columnwidth]{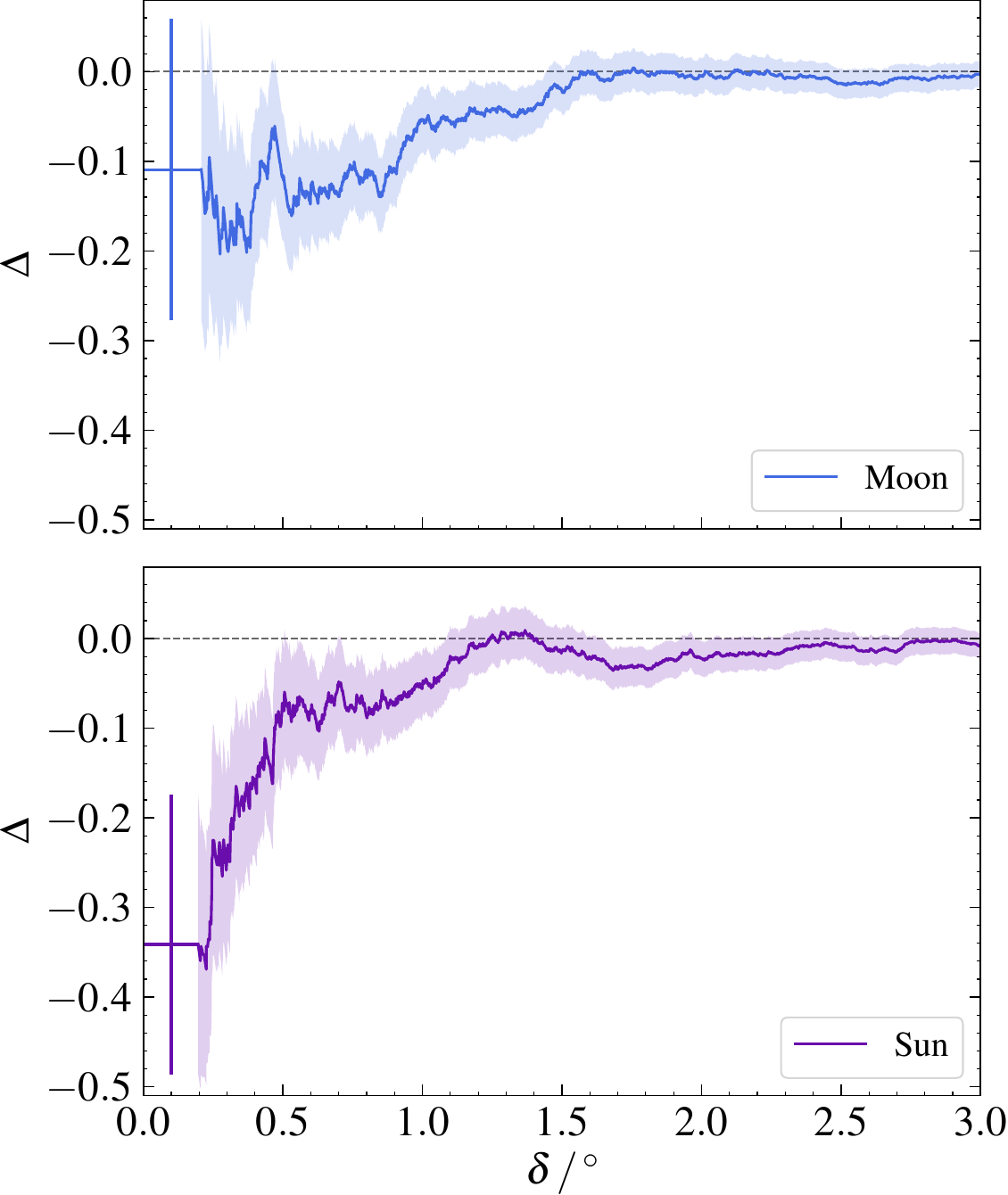}
\caption{Relative difference between the cumulative numbers of measured and expected events as a function of $\delta$ for the Moon (upper panel) and the Sun (lower panel).
The shaded bands represent the $1\sigma$ statistical uncertainties.
Because the relative difference is computed cumulatively, the data points are statistically correlated.}
\label{fig:cumul_moon_sun}
\end{figure}

The distributions for the Moon and the Sun are shown in \cref{fig:cumul_moon_sun} up to $\delta=\ang{3}$ for better visibility, since at larger distances they only converge to zero.
Deficits are observed at the centers of both celestial bodies, providing a verification of the pointing accuracy of the surface detector of the Observatory.

\begin{figure}
\includegraphics[width=\columnwidth]{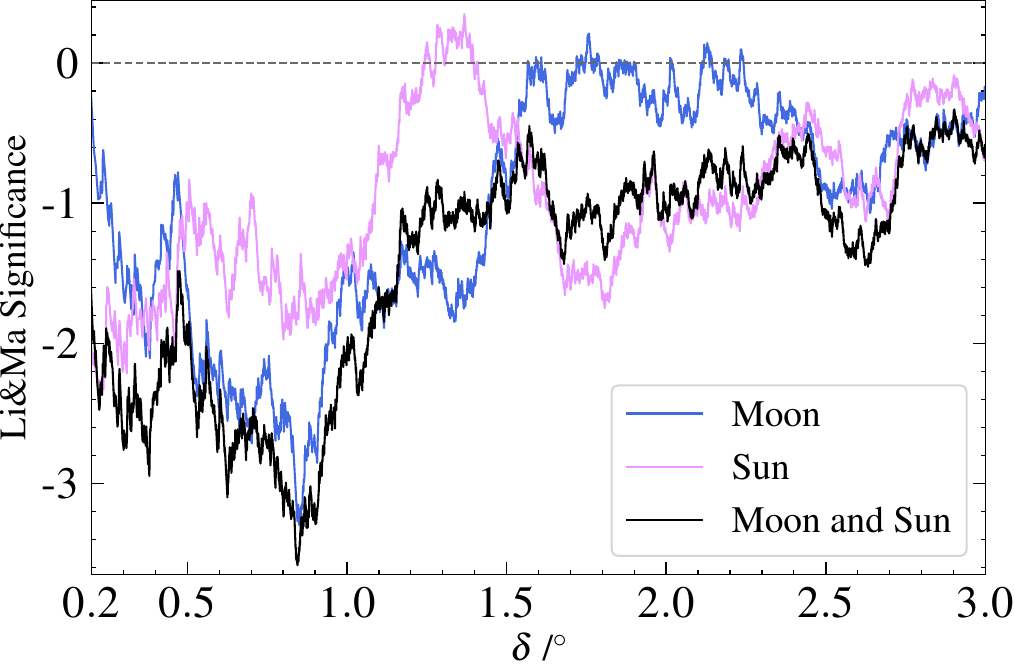}
\caption{Individual and combined (black) Li\&Ma significances near the Moon (blue) and the Sun (magenta) as functions of $\delta$.}
\label{fig:signif_moon_sun}
\end{figure}

The statistical significance of observing $N_\text{obs}$ events given the uniform expectation $N_\text{exp}$ is evaluated~\cite{gammapy:2023,acero_2025_14760974_gammapy} using the Li\&Ma formalism~\cite{LiMa}, with negligible uncertainty in the background.
The Li\&Ma significance as a function of the angular distance $\delta$ is shown in \cref{fig:signif_moon_sun}.

For the Moon, the Li\&Ma significance varies with $\delta$, increasing up to a maximum of $3.3\sigma$ at \ang{0.85}, while for the Sun it reaches a maximum of $2.4\sigma$ at \ang{0.23}.
The combined Li\&Ma significance is dominated by the Moon shadow, having a maximum of $3.6\sigma$ at $\delta=\ang{0.84}$. Although the deficit associated with the Sun appears more pronounced than that associated with the Moon in \cref{fig:cumul_moon_sun} (a difference with statistical significance below $1.5\sigma$), the Sun contributes less to the combined Li\&Ma significance owing  to the lower statistics at small angular distances.   

\section{Overall angular resolution}
\label{s:angular}

\begin{figure}
\includegraphics[width=\columnwidth]{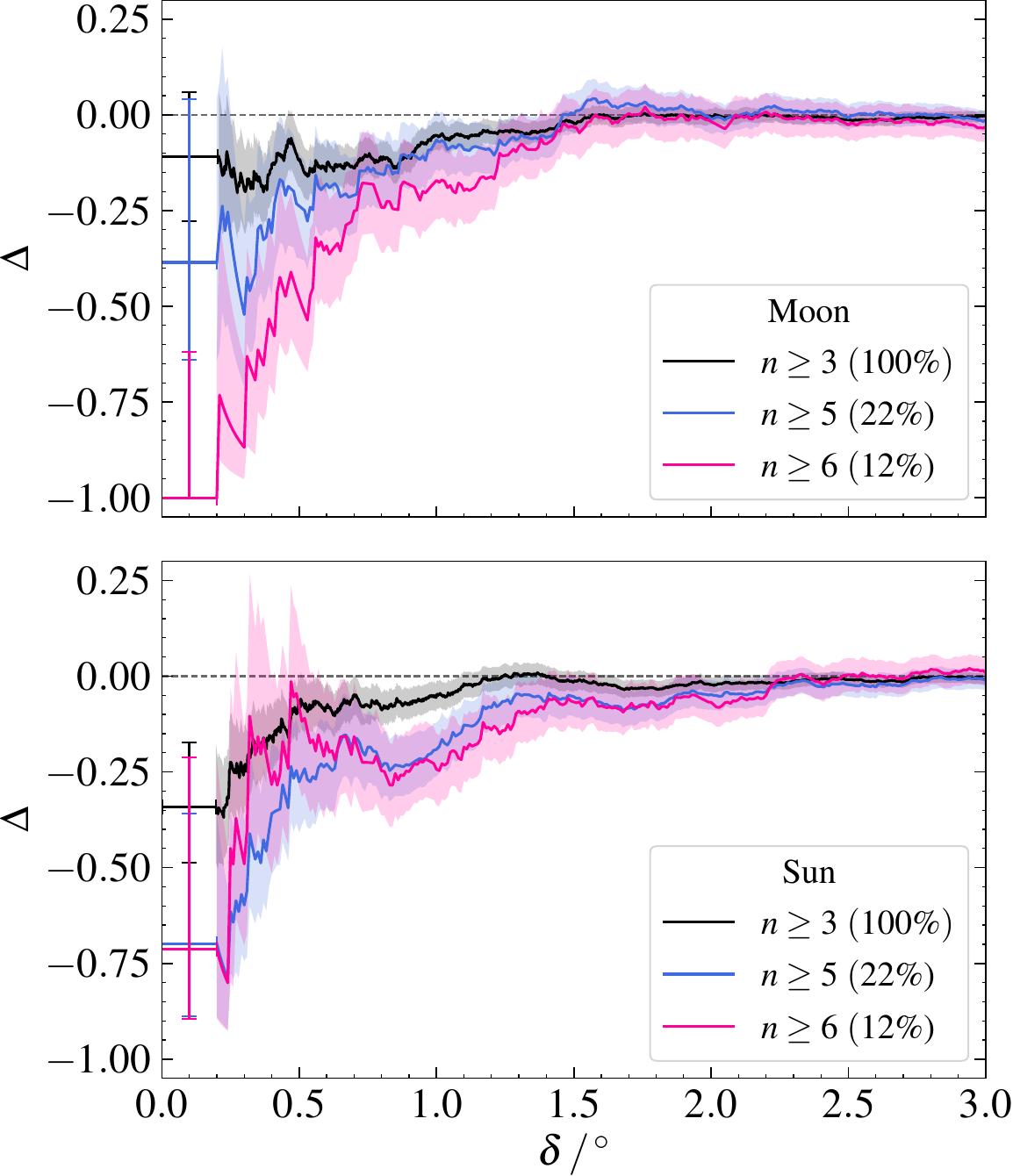}
\caption{Relative difference of the cumulative number of events as a function of the angular distance $\delta$ for event samples with different detector multiplicities $n$.
The fractions of each subsample relative to the full data set are indicated in the legend.}
\label{fig:cumul_Nstation}
\end{figure}

The angular resolution of the SD determines the shape of the shadows: better resolution leads to a deeper deficit at small angular distances $\delta$.
The angular resolution improves with the number of detectors participating in the event reconstruction, which in turn depends on the energy and the zenith angle of an event.
In \cref{fig:cumul_Nstation}, events are selected according to a minimum station multiplicity, ranging from the full data set with at least three detectors to subsamples with more than five detectors.
As expected, the deficit toward the center of the Moon and Sun becomes more pronounced with increasing multiplicity.
For the Moon, the relative deficit at $\delta=\ang{0.6}$ is $(13\pm6)\%$ for the full data set, $(18\pm12)\%$ for events reconstructed with more than four detectors, and it increases to $(34\pm15)\%$ for events with more than five detectors, showing the correct ordering.
The relative deficits for the Sun at $\delta=\ang{0.6}$ are $(8\pm6)\%$ for the full data set, $(23\pm11)\%$ with more than four detectors, and $(17\pm16)\%$ with more than five detectors, compatible with an ordering, however not conclusive due to the lower significance of the shadow of the Sun. 

To estimate the overall angular resolution of the three arrays of the SD, we employ an unbinned likelihood method similar to that described in Ref.~\cite{PhysRevD.49.1171_CASA}.
First, we consider that, without shadowing, the event count per unit solid angle is a constant $\left(\mathrm{d}N/\mathrm{d}\Omega\right)_\text{exp}=A$.
Next, we approximate the point spread function of SD to be a symmetric 2-D Gaussian of width $\sigma_\text{r}$ using a planar approximation over a small region of the sky.
Then the probability $p$ of the true cosmic-ray direction $\bf{\hat{x}}_\text{true}$ given the observed cosmic-ray direction $\bf{\hat{x}}_\text{obs}$ is
\begin{equation}
  p(\hat{\mathbf{x}}_\text{true} | \hat{\mathbf{x}}_\text{obs}, \sigma_\text{r}) =
    \frac{1}{2\pi\sigma_\text{r}^2}
    \exp\left[
      -\frac{\arccos(\hat{\mathbf{x}}_\text{true} \cdot \hat{\mathbf{x}}_\text{obs})^2}
            {2\sigma_\text{r}^2}
    \right].
\end{equation}
In a region centered on the Moon or Sun, the density of true directions is zero below the object's angular radius $\delta_\text{c}$ and $A$ above it, thus the expected density is
\begin{align}
\begin{split}
  A_{\sigma_\text{r}}(\hat{\mathbf{x}}_\text{obs}) &=
    A
    \int_{\delta>\delta_\text{c}}
      p(\hat{\mathbf{x}}_\text{true} | \hat{\mathbf{x}}_\text{obs}, \sigma_\text{r}) \,
      \mathrm{d}\Omega
\\
  &= A
  \left[
    1 -
    \int_{\delta<\delta_\text{c}}
      p(\hat{\mathbf{x}}_\text{true} | \hat{\mathbf{x}}_\text{obs}, \sigma_\text{r}) \,
      \mathrm{d}\Omega
  \right],
\end{split}
\end{align}
where the second integral is the probability that an arrival direction at $\hat{\mathbf{x}}_\text{obs}$ would have come from a direction within the solid angle of the Moon (or Sun), $\Omega$, if the body did not block cosmic rays.
The second integral is explicitly
\begin{align}
  \int^{\delta_\text{c}}_0 \!\!\!\!\! r'\,\mathrm{d}r'
    \int^{2\pi}_0 \!\!\!\!\! \mathrm{d}\phi' \,
      \frac{1}{2\pi\sigma_\text{r}^2}
      \exp\left[
        -\frac{\xi(\delta, r',\phi')^2}{2\sigma_\text{r}^2}
      \right],
\label{eq:like}
\end{align}
 where the integration is in polar coordinates $(r', \phi')$, with $r'$ ranging from zero to the angular radius of the celestial body.
The angle $\xi(\delta, r', \phi')$, equivalent to $\hat{\mathbf{x}}_\text{true} \cdot \hat{\mathbf{x}}_\text{obs}$, defines the angular distance between the point $(\delta, 0)$ at which the likelihood is evaluated, and the integration point $(r', \phi')$ on the disk, $\xi=[r'^2+\delta^2-2r'\delta \cos{\phi'}]^{1/2}$. Given the rotational symmetry of the problem, the likelihood is independent of the azimuthal angle and depends only on $\delta$.

This axisymmetric density function can be used as a probability distribution by a renormalization to give it a unit integral over the solid angle corresponding to $\delta\leq\ang{5}$,
\begin{align}
  \mathcal{P}(\delta) =
    \frac{A_{\sigma_\text{r}}(\delta)}{2\pi\int^{\ang{5}}_{0} A_{\sigma_\text{r}}\sin\delta\,\mathrm{d}\delta}
\end{align}

The cost function is then the sum of $\ln\mathcal{P}$ for each event, given one free parameter $\sigma_\mathrm{r}$.
Maximizing the likelihood yields $\sigma_\text{r} = (0.41^{+0.14}_{-0.09})\degree$ for the Moon data, $\sigma_\text{r} = (0.36^{+0.20}_{-0.10})\degree$ for the Sun data, and $\sigma_\text{r} = (0.39^{+0.10}_{-0.07})\degree$ for the combined Moon and Sun data.

The overall angular resolution enclosing 68\% of the events is calculated as $\sigma_{68} = \sqrt{2.278}\,\sigma_\text{r}$.
This corresponds to $\sigma_{68} = (0.61^{+0.21}_{-0.13})\degree$ for the Moon data, $\sigma_{68} = (0.54^{+0.30}_{-0.15})\degree$ for the Sun data, and $\sigma_{68} = (0.59^{+0.15}_{-0.11})\degree$ for the combined data.

The events in the data sample do not share a single angular resolution; instead, $\sigma_{68}$ reflects a combination of the resolution dependencies on zenith angle and station multiplicity, consistent with the multiplicity-dependent behavior discussed above.
A Monte-Carlo study using event-by-event resolution estimates from the reconstruction was performed to verify the consistency of the fitted $\sigma_\text{r}$ with existing measurements of the angular resolution of SD.
The most probable value obtained from the simulation is consistent with the result presented here; further details are provided in the \cref{app:toyMC}.

The significance of the parameter estimate is evaluated using the likelihood ratio test~\cite{PhysRevD.43.1735_Cygnus},
\begin{align}
  N_\sigma = \sqrt{2(w_\text{max}-w_\infty)},
\label{eq:fit_signif}
\end{align}
where $w_\text{max}$ is the log-likelihood at the best-fit resolution value and $w_\infty$ is the log-likelihood corresponding to a very poor resolution (any angle greater than \ang{3} in this case).
For the Moon data, the fit yields $N_\sigma = 2.39$, while for the Sun data $N_\sigma = 1.78$.
For the combined Moon and Sun data, the significance increases to $N_\sigma = 2.97$.

Given the estimated overall angular resolution from the combined shadows, the maximum deficit significance is expected at $\delta = (0.65^{+0.16}_{-0.11})\degree$.
Comparison with the Li\&Ma significance shown in \cref{fig:signif_moon_sun} indicates that the observed maximum is within $1.2\sigma$ consistent with this interval.
Based on the deficit of events measured at $\delta = \ang{0.65}$ and that according to our likelihood model, we expect 71\% of the total deficit to be within this radius, we can calculate the total observed deficit.
Near the Moon, it is $57.7 \pm 23.1$ events, compared to an expected deficit of 49.4 events derived from the uniform background and the solid angle of the Moon.
Near the Sun, the observed deficit is $38.1 \pm 24.0$ events, while the corresponding expectation is 50.6 events.

The largest systematic uncertainty on the angular resolution, $\pm\ang{0.03}$, arises from the assumption of a fixed angular radius for the Moon and the Sun.
This uncertainty is estimated by varying the angular radii between their minimum and maximum apparent values: \ang{0.245} to \ang{0.279} for the Moon and \ang{0.262} to \ang{0.272} for the Sun.

Additional effects are negligible, including those arising from the choice of the coordinate-system origin over the \qty{3000}{km^2} extent of the observatory entering the coordinate transformation, deflections of cosmic-ray trajectories by the geomagnetic field (of order ${\sim}\ang{0.01}$ for iron nuclei at energies of \qty{10}{PeV}~\cite{Aartsen_2014_icecube}), and photodisintegration of cosmic rays by sunlight~\cite{Andersen:2011dz}.
Deviations induced by the solar magnetic field at energies above \qty{e16}{eV} are also expected to be negligible~\cite{CRpropagSun}.

\section{Conclusion}
\label{s:conclusion}

More than ten million events recorded by the Pierre Auger Observatory were analyzed to study the local deficit in the flux of cosmic rays above \qty{e16}{eV} caused by the presence of the Moon and the Sun.
This event sample was obtained by combining data from the three arrays of the Observatory: SD-1500, SD-750, and SD-433.
The data can be combined based on the similarities in the detector responses and reconstructions.

A one-dimensional analysis of the angular distance between the cosmic-ray arrival directions and the centers of the Moon and Sun reveals a statistically significant deficit of approximately $3\sigma$.

From the measured shadows of the Moon and the Sun, we determine the overall angular resolution of the SD to be $\sigma_{68}=(0.59^{+0.15}_{-0.11})\degree$.
Moreover, this result provides an independent validation of the pointing accuracy and angular resolution of the SD of the Pierre Auger Observatory at a mean cosmic-ray energy of \qty{7e17}{eV}. The number of shadowed events is consistent with the expected shadowing based on the size of the celestial bodies and the obtained overall angular resolution.

The Pierre Auger Observatory will continue its operations until at least the year 2035, and the accumulated event statistics will increase the significance of these observations.
It will allow for a more precise determination of the overall angular resolution using the Moon and Sun shadows.

\begin{acknowledgments}
This work is dedicated to the memory of Jim Cronin, Nobel laureate, co-founder of the Pierre Auger Collaboration, and its long-term spokesperson, whose seminal contributions to cosmic-ray physics include a similar study using data from the Chicago Air Shower Array (CASA)~\cite{PhysRevD.49.1171_CASA}.


\begin{sloppypar}
The successful installation, commissioning, and operation of the Pierre
Auger Observatory would not have been possible without the strong
commitment and effort from the technical and administrative staff in
Malarg\"ue. We are very grateful to the following agencies and
organizations for financial support:
\end{sloppypar}

\begin{sloppypar}
Argentina -- Comisi\'on Nacional de Energ\'\i{}a At\'omica; Agencia Nacional de
Promoci\'on Cient\'\i{}fica y Tecnol\'ogica (ANPCyT); Consejo Nacional de
Investigaciones Cient\'\i{}ficas y T\'ecnicas (CONICET); Gobierno de la
Provincia de Mendoza; Municipalidad de Malarg\"ue; NDM Holdings and Valle
Las Le\~nas; in gratitude for their continuing cooperation over land
access; Australia -- the Australian Research Council; Belgium -- Fonds
de la Recherche Scientifique (FNRS); Research Foundation Flanders (FWO),
Marie Curie Action of the European Union Grant No.~101107047; Brazil --
Minist\'erio da Ci\^encia, Tecnologia e Inova\c{c}\~ao (MCTI); Czech Republic --
GACR 24-13049S, CAS LQ100102401, MEYS LM2023032,
CZ.02.1.01/0.0/0.0/16{\textunderscore}013/0001402, CZ.02.1.01/0.0/0.0/18{\textunderscore}046/0016010
and CZ.02.1.01/0.0/0.0/17{\textunderscore}049/0008422 and
CZ.02.01.01/00/22{\textunderscore}008/0004632; France -- Centre de Calcul IN2P3/CNRS;
Centre National de la Recherche Scientifique (CNRS); Institut National
de Physique Nucl\'eaire et de Physique des Particules (IN2P3/CNRS);
Germany -- Bundesministerium f\"ur Forschung, Technologie und Raumfahrt
(BMFTR); Deutsche Forschungsgemeinschaft (DFG); Ministerium f\"ur Finanzen
Baden-W\"urttemberg; Helmholtz Alliance for Astroparticle Physics (HAP);
Hermann von Helmholtz-Gemeinschaft Deutscher Forschungszentren e.V.;
Ministerium f\"ur Kultur und Wissenschaft des Landes Nordrhein-Westfalen;
Ministerium f\"ur Wissenschaft, Forschung und Kunst des Landes
Baden-W\"urttemberg; Italy -- Istituto Nazionale di Fisica Nucleare
(INFN); Istituto Nazionale di Astrofisica (INAF); Ministero
dell'Universit\`a e della Ricerca (MUR); CETEMPS Center of Excellence;
Ministero degli Affari Esteri (MAE), ICSC Centro Nazionale di Ricerca in
High Performance Computing, Big Data and Quantum Computing, funded by
European Union NextGenerationEU, reference code CN{\textunderscore}00000013; M\'exico --
Consejo Nacional de Ciencia y Tecnolog\'\i{}a (CONACYT-SECHTI)
No.~CB-A1-S-46703, Universidad Nacional Aut\'onoma de M\'exico (UNAM)
PAPIIT-IN114924; Benem\'erita Universidad Aut\'onoma de Puebla (BUAP), VIEP
and Laboratorio Nacional de Superc\'omputo del Sureste de M\'exico (LNS);
and Benem\'erita Universidad Aut\'onoma de Chiapas (UNACH); The Netherlands
-- Ministry of Education, Culture and Science; Netherlands Organisation
for Scientific Research (NWO); Dutch national e-infrastructure with the
support of SURF Cooperative; Poland -- Ministry of Science and Higher
Education, grant No.~2022/WK/12; National Science Centre, grants
No.~2020/39/B/ST9/01398, and 2022/45/B/ST9/02163; Portugal -- Portuguese
national funds and FEDER funds within Programa Operacional Factores de
Competitividade through Funda\c{c}\~ao para a Ci\^encia e a Tecnologia
(COMPETE); Romania -- Ministry of Education and Research, contract
no.~30N/2023 under Romanian National Core Program LAPLAS VII, and grant
no.~PN 23 21 01 02; Slovenia -- Slovenian Research and Innovation
Agency, grants P1-0031, I0-0033; Spain -- Ministerio de Ciencia,
Innovaci\'on y Universidades/Agencia Estatal de Investigaci\'on MICIU/AEI
/10.13039/501100011033 (PID2022-140510NB-I00, PCI2023-145952-2,
CNS2024-154676, and Mar\'\i{}a de Maeztu CEX2023-001318-M), Xunta de Galicia
(CIGUS Network of Research Centers, Consolidaci\'on ED431C-2025/11 and
ED431F-2022/15) and European Union ERDF; USA -- Department of Energy,
Contracts No.~DE-AC02-07CH11359, No.~DE-FR02-04ER41300,
No.~DE-FG02-99ER41107 and No.~DE-SC0011689; National Science Foundation,
Grant No.~0450696, and NSF-2013199; The Grainger Foundation;
Astrophysics Centre for Multi-messenger studies in Europe (ACME) EU
Grant No 101131928; and UNESCO.
\end{sloppypar}

\end{acknowledgments}

\appendix
\refstepcounter{appendixref}

\section*{Appendix}
\label{app:toyMC}

The obtained overall angular resolution, $\sigma_{68}$, is a combination of multiple event-by-event resolutions for events with different station multiplicities and different station separations. Therefore, its interpretation is not straightforward.

To compare our result with previous resolution studies~\cite{Aab_2020_Auger_reco, neutron_icrc23,AMIGA_2011,Silli:2021Jt_SD-433}, simulation samples were generated using a time-shuffling technique applied on the events, which varied their right ascensions.
This resulted in a dataset of isotropic events approximately 200 times larger than the original event sample.
For each simulation sample, a subset of the events within \ang{10} of the fake celestial bodies was randomly selected to match the size of the data sample under study.
The shadows were introduced by removing events within \ang{0.26} of the center of the fake celestial bodies. 

The arrival directions of the remaining events were then randomized based on the reconstruction uncertainties characteristic of each event (depending on the station multiplicity, $n$, and SD array).
The 68\% quantiles of the resolution distributions used are $\sigma_{n=3}=\ang{1.71}$, $\sigma_{n=4}=\ang{1.19}$, and $\sigma_{n\geq5}=\ang{0.75}$ for the SD-1500, and $\sigma_{n=3}=\ang{1.28}$, $\sigma_{n=4}=\ang{1.04}$, and $\sigma_{n\geq5}=\ang{0.89}$ for the SD-750.
The SD-433 events were not included in these Monte-Carlo tests as they represent just 6\% of the total data sample.

\begin{figure}[t]
\includegraphics[width=\columnwidth]{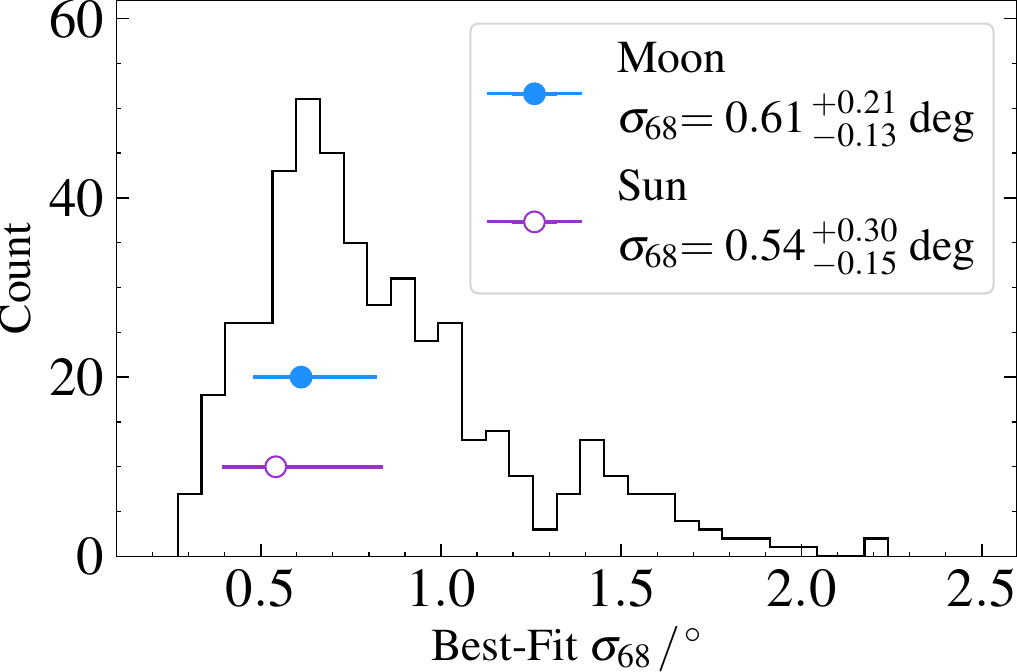}
\caption{The distribution of the $\sigma_\text{68}$ values obtained from the Monte-Carlo simulations.
The result of the Moon-shadow fit is indicated by a filled blue marker, while the result of the Sun-shadow fit is shown with an open purple circle.}
\label{fig:app-toyMC}
\end{figure}

The overall angular resolution is obtained for each simulated sample using an identical method to the one described in \cref{s:angular}.
The distribution of the obtained best-fit values is shown in \cref{fig:app-toyMC}.
The most probable value (mode) is $0.65\pm\ang{0.02}$, compatible with the results obtained from the Moon and Sun shadow fits, shown with blue and purple markers in the same figure.

\balance
\clearpage

\bibliography{my_bib}


\clearpage

\onecolumngrid

\begin{center}
\rule{0.1\columnwidth}{0.5pt}
\raisebox{-0.4ex}{\scriptsize$\bullet$}
\rule{0.1\columnwidth}{0.5pt}
\end{center}
\vspace{3mm}

\begin{wrapfigure}[8]{l}{0.06\textwidth}
\vspace{-2.9ex}
\includegraphics[width=1.4\linewidth]{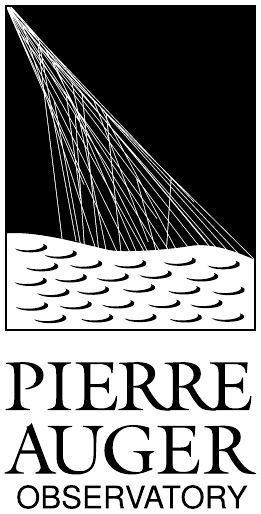}
\end{wrapfigure}
\begin{sloppypar}\noindent
A.~Abdul Halim$^{13}$,
P.~Abreu$^{67}$,
M.~Aglietta$^{50,49}$,
M.~Ahmed$^{31}$,
I.~Allekotte$^{1}$,
K.~Almeida Cheminant$^{74,73}$,
R.~Aloisio$^{42,43}$,
J.~Alvarez-Mu\~niz$^{72}$,
A.~Ambrosone$^{42,43}$,
J.~Ammerman Yebra$^{72}$
% ...

\end{sloppypar}
\begin{center}
\par\noindent
\textbf{The Pierre Auger Collaboration}
\end{center}

\vspace{1ex}
\begin{description}[labelsep=0.2em,align=right,labelwidth=0.7em,labelindent=0em,leftmargin=2em,noitemsep,before={\renewcommand\makelabel[1]{##1 }}]
\item[$^{1}$] Centro At\'omico Bariloche and Instituto Balseiro (CNEA-UNCuyo-CONICET), San Carlos de Bariloche, Argentina
\item[$^{2}$] Departamento de F\'\i{}sica and Departamento de Ciencias de la Atm\'osfera y los Oc\'eanos, FCEyN, Universidad de Buenos Aires and CONICET, Buenos Aires, Argentina
\item[$^{3}$] IFLP, Universidad Nacional de La Plata and CONICET, La Plata, Argentina
\item[$^{4}$] Instituto de Astronom\'\i{}a y F\'\i{}sica del Espacio (IAFE, CONICET-UBA), Buenos Aires, Argentina
\item[$^{5}$] Instituto de F\'\i{}sica de Rosario (IFIR) -- CONICET/U.N.R.\ and Facultad de Ciencias Bioqu\'\i{}micas y Farmac\'euticas U.N.R., Rosario, Argentina
\item[$^{6}$] Instituto de Tecnolog\'\i{}as en Detecci\'on y Astropart\'\i{}culas (CNEA, CONICET, UNSAM), and Universidad Tecnol\'ogica Nacional -- Facultad Regional Mendoza (CONICET/CNEA), Mendoza, Argentina
\item[$^{7}$] Instituto de Tecnolog\'\i{}as en Detecci\'on y Astropart\'\i{}culas (CNEA, CONICET, UNSAM), Buenos Aires, Argentina
\item[$^{8}$] International Center of Advanced Studies and Instituto de Ciencias F\'\i{}sicas, ECyT-UNSAM and CONICET, Campus Miguelete -- San Mart\'\i{}n, Buenos Aires, Argentina
\item[$^{9}$] Laboratorio Atm\'osfera -- Departamento de Investigaciones en L\'aseres y sus Aplicaciones -- UNIDEF (CITEDEF-CONICET), Argentina
\item[$^{10}$] Observatorio Pierre Auger, Malarg\"ue, Argentina
\item[$^{11}$] Observatorio Pierre Auger and Comisi\'on Nacional de Energ\'\i{}a At\'omica, Malarg\"ue, Argentina
\item[$^{12}$] Universidad Tecnol\'ogica Nacional -- Facultad Regional Buenos Aires, Buenos Aires, Argentina
\item[$^{13}$] Adelaide University, Adelaide, S.A., Australia
\item[$^{14}$] Universit\'e Libre de Bruxelles (ULB), Brussels, Belgium
\item[$^{15}$] Vrije Universiteit Brussels, Brussels, Belgium
\item[$^{16}$] Centro Brasileiro de Pesquisas Fisicas, Rio de Janeiro, RJ, Brazil
\item[$^{17}$] Centro Federal de Educa\c{c}\~ao Tecnol\'ogica Celso Suckow da Fonseca, Petropolis, Brazil
\item[$^{18}$] Universidade de S\~ao Paulo, Escola de Engenharia de Lorena, Lorena, SP, Brazil
\item[$^{19}$] Universidade de S\~ao Paulo, Instituto de F\'\i{}sica de S\~ao Carlos, S\~ao Carlos, SP, Brazil
\item[$^{20}$] Universidade de S\~ao Paulo, Instituto de F\'\i{}sica, S\~ao Paulo, SP, Brazil
\item[$^{21}$] Universidade Estadual de Campinas (UNICAMP), IFGW, Campinas, SP, Brazil
\item[$^{22}$] Universidade Estadual de Feira de Santana, Feira de Santana, Brazil
\item[$^{23}$] Universidade Federal do ABC, Santo Andr\'e, SP, Brazil
\item[$^{24}$] Universidade Federal do Paran\'a, Setor Palotina, Palotina, Brazil
\item[$^{25}$] Universidade Federal do Rio de Janeiro, Instituto de F\'\i{}sica, Rio de Janeiro, RJ, Brazil
\item[$^{26}$] Universidad de Medell\'\i{}n, Medell\'\i{}n, Colombia
\item[$^{27}$] Universidad Industrial de Santander, Bucaramanga, Colombia
\item[$^{28}$] Charles University, Faculty of Mathematics and Physics, Institute of Particle and Nuclear Physics, Prague, Czech Republic
\item[$^{29}$] Institute of Physics of the Czech Academy of Sciences, Prague, Czech Republic
\item[$^{30}$] Palacky University, Olomouc, Czech Republic
\item[$^{31}$] CNRS/IN2P3, IJCLab, Universit\'e Paris-Saclay, Orsay, France
\item[$^{32}$] Laboratoire de Physique Nucl\'eaire et de Hautes Energies (LPNHE), Sorbonne Universit\'e, Universit\'e de Paris, CNRS-IN2P3, Paris, France
\item[$^{33}$] Universit\'e Paris-Saclay, CNRS/IN2P3, IJCLab, Orsay, France
\item[$^{34}$] Bergische Universit\"at Wuppertal, Department of Physics, Wuppertal, Germany
\item[$^{35}$] Karlsruhe Institute of Technology (KIT), Institute for Experimental Particle Physics, Karlsruhe, Germany
\item[$^{36}$] Karlsruhe Institute of Technology (KIT), Institut f\"ur Prozessdatenverarbeitung und Elektronik, Karlsruhe, Germany
\item[$^{37}$] Karlsruhe Institute of Technology (KIT), Institute for Astroparticle Physics, Karlsruhe, Germany
\item[$^{38}$] RWTH Aachen University, III.\ Physikalisches Institut A, Aachen, Germany
\item[$^{39}$] TU Dortmund University, Department of Physics, Dortmund, Germany
\item[$^{40}$] Universit\"at Hamburg, II.\ Institut f\"ur Theoretische Physik, Hamburg, Germany
\item[$^{41}$] Universit\"at Siegen, Department Physik -- Experimentelle Teilchenphysik, Siegen, Germany
\item[$^{42}$] Gran Sasso Science Institute, L'Aquila, Italy
\item[$^{43}$] INFN Laboratori Nazionali del Gran Sasso, Assergi (L'Aquila), Italy
\item[$^{44}$] INFN, Sezione di Catania, Catania, Italy
\item[$^{45}$] INFN, Sezione di Lecce, Lecce, Italy
\item[$^{46}$] INFN, Sezione di Milano, Milano, Italy
\item[$^{47}$] INFN, Sezione di Napoli, Napoli, Italy
\item[$^{48}$] INFN, Sezione di Roma ``Tor Vergata'', Roma, Italy
\item[$^{49}$] INFN, Sezione di Torino, Torino, Italy
\item[$^{50}$] Osservatorio Astrofisico di Torino (INAF), Torino, Italy
\item[$^{51}$] Politecnico di Milano, Dipartimento di Scienze e Tecnologie Aerospaziali , Milano, Italy
\item[$^{52}$] Universit\`a del Salento, Dipartimento di Matematica e Fisica ``E.\ De Giorgi'', Lecce, Italy
\item[$^{53}$] Universit\`a dell'Aquila, Dipartimento di Scienze Fisiche e Chimiche, L'Aquila, Italy
\item[$^{54}$] Universit\`a di Catania, Dipartimento di Fisica e Astronomia ``Ettore Majorana``, Catania, Italy
\item[$^{55}$] Universit\`a di Milano, Dipartimento di Fisica, Milano, Italy
\item[$^{56}$] Universit\`a di Napoli ``Federico II'', Dipartimento di Fisica ``Ettore Pancini'', Napoli, Italy
\item[$^{57}$] Universit\`a di Palermo, Dipartimento di Fisica e Chimica ''E.\ Segr\`e'', Palermo, Italy
\item[$^{58}$] Universit\`a di Roma ``Tor Vergata'', Dipartimento di Fisica, Roma, Italy
\item[$^{59}$] Universit\`a Torino, Dipartimento di Fisica, Torino, Italy
\item[$^{60}$] Benem\'erita Universidad Aut\'onoma de Puebla, Puebla, M\'exico
\item[$^{61}$] Unidad Profesional Interdisciplinaria en Ingenier\'\i{}a y Tecnolog\'\i{}as Avanzadas del Instituto Polit\'ecnico Nacional (UPIITA-IPN), M\'exico, D.F., M\'exico
\item[$^{62}$] Universidad Aut\'onoma de Chiapas, Tuxtla Guti\'errez, Chiapas, M\'exico
\item[$^{63}$] Universidad Michoacana de San Nicol\'as de Hidalgo, Morelia, Michoac\'an, M\'exico
\item[$^{64}$] Universidad Nacional Aut\'onoma de M\'exico, M\'exico, D.F., M\'exico
\item[$^{65}$] Institute of Nuclear Physics PAN, Krakow, Poland
\item[$^{66}$] University of \L{}\'od\'z, Faculty of High-Energy Astrophysics,\L{}\'od\'z, Poland
\item[$^{67}$] Laborat\'orio de Instrumenta\c{c}\~ao e F\'\i{}sica Experimental de Part\'\i{}culas -- LIP and Instituto Superior T\'ecnico -- IST, Universidade de Lisboa -- UL, Lisboa, Portugal
\item[$^{68}$] ``Horia Hulubei'' National Institute for Physics and Nuclear Engineering, Bucharest-Magurele, Romania
\item[$^{69}$] Institute of Space Science, Bucharest-Magurele, Romania
\item[$^{70}$] Center for Astrophysics and Cosmology (CAC), University of Nova Gorica, Nova Gorica, Slovenia
\item[$^{71}$] Experimental Particle Physics Department, J.\ Stefan Institute, Ljubljana, Slovenia
\item[$^{72}$] Instituto Galego de F\'\i{}sica de Altas Enerx\'\i{}as (IGFAE), Universidade de Santiago de Compostela, Santiago de Compostela, Spain
\item[$^{73}$] IMAPP, Radboud University Nijmegen, Nijmegen, The Netherlands
\item[$^{74}$] Nationaal Instituut voor Kernfysica en Hoge Energie Fysica (NIKHEF), Science Park, Amsterdam, The Netherlands
\item[$^{75}$] Stichting Astronomisch Onderzoek in Nederland (ASTRON), Dwingeloo, The Netherlands
\item[$^{76}$] Case Western Reserve University, Cleveland, OH, USA
\item[$^{77}$] Colorado School of Mines, Golden, CO, USA
\item[$^{78}$] Department of Physics and Astronomy, Lehman College, City University of New York, Bronx, NY, USA
\item[$^{79}$] Michigan Technological University, Houghton, MI, USA
\item[$^{80}$] New York University, New York, NY, USA
\item[$^{81}$] University of Chicago, Enrico Fermi Institute, Chicago, IL, USA
\item[$^{82}$] University of Delaware, Department of Physics and Astronomy, Bartol Research Institute, Newark, DE, USA
\item[] -----
\item[$^{a}$] Max-Planck-Institut f\"ur Radioastronomie, Bonn, Germany
\item[$^{b}$] also at Kapteyn Institute, University of Groningen, Groningen, The Netherlands
\item[$^{c}$] School of Physics and Astronomy, University of Leeds, Leeds, United Kingdom
\item[$^{d}$] Fermi National Accelerator Laboratory, Fermilab, Batavia, IL, USA (Affiliation for identification purposes only)
\item[$^{e}$] Pennsylvania State University, University Park, PA, USA
\item[$^{f}$] Colorado State University, Fort Collins, CO, USA
\item[$^{g}$] Louisiana State University, Baton Rouge, LA, USA
\item[$^{h}$] Universit\'e Grenoble Alpes, CNRS, Grenoble Institute of Engineering, LPSC-IN2P3, Grenoble, France
\item[$^{i}$] Institut universitaire de France (IUF), France
\item[$^{j}$] now at Graduate School of Science, Osaka Metropolitan University, Osaka, Japan
\item[$^{k}$] now at Lek (Sandoz), Ljubljana, Slovenia
\end{description}


\end{document}